\documentclass[%
 reprint,superscriptaddress,amsmath,amssymb,prb,showpacs]{revtex4-1}
\usepackage{graphicx}
\usepackage{dcolumn}
\usepackage{bm}
\usepackage{hyperref}

\begin{document}
	\title{Cluster-correlation expansion for studying decoherence of clock transitions in spin baths}
	\author{Geng-Li Zhang}
	\affiliation{Department of Physics and The Hong Kong Institute of Quantum Information of Science and Technology, The Chinese University of Hong Kong, Shatin, N. T., Hong Kong, China}
	
	\author{Wen-Long Ma}
	\altaffiliation[Current Address: ]{Pritzker School of Molecular Engineering, University of Chicago, Illinois 60637, USA}
	\affiliation{Department of Physics and The Hong Kong Institute of Quantum Information of Science and Technology, The Chinese University of Hong Kong, Shatin, N. T., Hong Kong, China}
	
	\author{Ren-Bao Liu}
	\email{rbliu@cuhk.edu.hk}
	\affiliation{Department of Physics and The Hong Kong Institute of Quantum Information of Science and Technology, The Chinese University of Hong Kong, Shatin, N. T., Hong Kong, China}
	\affiliation{Institute of Theoretical Physics, The Chinese University of Hong Kong, Shatin, N.T., Hong Kong, China}

	\date{\today }
	
	\begin{abstract}
	The clock transitions (CTs) of central spins have long coherence times because their frequency fluctuations vanish in the linear order of external field noise (such as Overhauser fields from nuclear spin baths). Therefore, CTs are useful for quantum technologies. Also, the quadratic dependence of frequencies on noises makes the CT decoherence an interesting physics problem. Thus we are motivated to study the decoherence of CTs. We consider noise from spin baths, which is one of the most relevant mechanisms of qubit decoherence. Various quantum many-body methods have been developed to study the decoherence of a central spin in spin baths. In particular, the cluster-correlation expansion (CCE) systematically accounts for the many-body correlations that cause the central spin decoherence. However, the CCE can not be straightforwardly applied to CTs in spin baths, for the expansion may fail to converge due to the effective long-range interactions resulting from the quadratic term of the noise (e.g.,  the second-order interaction mediated by hyperfine interactions for a nuclear spin bath). In this work, we develop a modified CCE method to tackle this class of decoherence problems. By diagonalizing the central spin Hamiltonian for each bath eigenstate of the hyperfine interaction, we find that the effects of long-range interactions are absorbed as fluctuations of central spin eigenenergies in the form of single-spin correlations. We apply the method to two specific systems, namely, nitrogen vacancy center electron spins in near zero magnetic field and singlet-triplet transition of two electrons in a double quantum dot. The numerical simulation shows that the modified CCE converges rapidly for the CTs. 
	\end{abstract}
	\maketitle
	
	\section{Introduction}
	Central spin decoherence due to coupling to environments \cite{Zurek2003, Schlosshauer2005, Prokofev2000} sets a limit for spin-based quantum techonogies \cite{Childress2006, T.D.Ladd2010, Rondin2014, Schirhagl2014}. Therefore, it is important to study the decoherence mechanism and to develop methods to prolong the coherence time \cite{Zhao2012}. In light-element materials (e.g., diamond) with weak spin-orbit interaction or at low temperatures, the phonon-induced qubit decoherence is negligible and the nuclear spin fluctuations become a dominant decoherence mechanism. To quantitatively describe the decoherence of electron spins in nuclear spin baths, quantum many-body approaches have been developed, including pair-correlation approximation \cite{Yao2006, Liu2007, Yao2007}, cluster expansion (CE) \cite{Witzel2005, Witzel2006}, linked-cluster expansion (LCE) \cite{Saikin2007}, cluster-correlation expansion (CCE) \cite{Yang2008, Yang2009} and ring diagram approximation \cite{Cywifmmodecutenlseniski2009, Cywifmmodecutenlseniski2009a}. 
	
	Among these theories, CCE is particularly useful for spin baths with random qubit-bath couplings. It has been successfully applied to various systems such as the nitrogen vacancy (NV) centers in diamond \cite{Gruber1997} and defect centers in silicon \cite{Morton2011}. The CCE approximates the coherence as factorized expansions, $L(t)\approx \prod_{|C|\leq K} \tilde{L}_C$, with $|C|$ being the number of spins contained in the cluster $C$ and $K$ being the truncation order \cite{Yang2008}. Typically CCE converges at small $K$ , and therefore is very efficient in simulations. 
	
	However, similar to other cluster expansion theories, CCE often does not work in the regime near clock transitions (CTs). At CTs the transition frequency of the central spin vanishes in the linear order of external field, i.e., $df/dB=0$ \cite{Wolfowicz2013a, Shiddiq2016} .The CTs are of interest to quantum technologies because of their inherently insensitivity to noise fluctuations. CT often occurs at level anti-crossing point. For example, in an NV center \cite{Gruber1997, Kennedy2003, Gaebel2006} under near zero field, the strain mixes two Zeeman energy states, leading to a parabolic dependence of the energy splitting on magnetic field and hence eliminating the effect of magnetic noises at the linear order \cite{Jamonneau2016}. The second order effect of  the Overhauser field due to hyperfine interactions results in an electron spin mediated long-range interaction between the nuclear spins (terms $\propto {\mathbf I}_i \cdot {\mathbf A}_i{\mathbf A}_j\cdot {\mathbf I}_j$).	Because of the long-range interactions, straightforward CCE often does not converge. 
	
	In this paper, we develop a modified CCE method to calculate electron spin decoherence caused by nuclear spin baths near CTs. We demonstrate the method by studying the decoherence of an NV center spin under near zero field with finite strain. Using the eigenstates of the hyperfine interaction as the basis of the nuclear spins and diagonalizing the electron spin Hamiltonian including the hyperfine energy, we transform the effects of the mediated long-range interactions between nuclear spins into the energy fluctuation of the electron spin eigenstates. The intrinsic dipolar interaction between nulear spins is treated by 2nd- and higher-order cluster expansion. Thus, for each initial nuclear spin state (i.e., each sampling of the Overhauser field), the CCE converges rapidly.  Actually, the dominating effects of the nuclear spin baths comes from the first order CCE, i.e., CCE-1 or single nuclear spin dynamics. The final result of decoherence is obtained by ensemble average over the thermal distribution of the Overhauser field.
	
	The modified CCE method can be applied to a broad class of other important scenarios including the singlet-triplet decoherence in GaAs double quantum dot \cite{Yang2008a, Jacak1998, Hanson2007}, decay of Rabi oscillations \cite{Koppens2007, Dobrovitski2009} and rotary echos \cite{Laraoui2011}. As a demonstration we also calculate the singlet-triplet decoherence in a double quantum dot. 
	
	This paper is organized as follows. Sec.~\ref{sec: Model} describes a general class of decoherence problems that can be solved by the modified CCE method. 
	As an example, we study in details the NV center in near zero magnetic field and show that the modified CCE converges quickly in Sec.~\ref{sec: NV}. The conclusion is given in Sec.~\ref{sec: conclusion}. In Appendix~\ref{sec: StaticBasis} we show that the electron spin eigenbasis can be approximated as static in cases of interest. The investigation of the singlet-triplet decoherence in GaAs double quantum dot is given in Appendix~\ref{Sec: Double-QDs}. In Appendix~\ref{sec: RotaryEcho} we show that the decay of Rabi oscillations and rotary echos in spin baths can also be studied with the modified CCE. 
	
	\section{Model and formalism}
	\label{sec: Model}
	We consider a central spin coupled to a spin bath. The Hamiltonian is
	\begin{equation}\label{eq: orgHam}
	H = \frac12 \Delta \sigma_z + \frac12 \varepsilon\sigma_x + H_\mathrm{b},
	\end{equation}
	where $\Delta$ is an effective field in the $z$ direction (e.g., external magnetic field and Overhauser field due to hyperfine coupling to nuclear spins), $\varepsilon$ is an effective field in the $x$ direction (e.g., strain), and $H_\mathrm{b}$ is the bath Hamiltonian.  
	
	When the bath interaction is neglected for the moment, the rest part of the Hamiltonian can be readily diagonalized, with the bath in an eigenstate $|M \rangle$ of the Overhauser field with eigenvalue $\Delta_M$, and the central spin diagonalized as 
	\begin{equation}\label{eq: GenHam}
	H_M = \frac 12 \sqrt{\varepsilon^2 + \Delta_M^2}\sigma_{\tilde{z}}^M,
	\end{equation}
	where $\sigma_{\tilde{z}}^M = \lvert + \rangle\langle + \rvert - \lvert -\rangle\langle-\rvert$ is the Pauli matrix in a basis depending on the bath state $\lvert M \rangle$. Therefore we treat the eigenenergy $E_{M} = \sqrt{\varepsilon^2 + \Delta_M^2}$ exactly rather than expand it to the second order (an approximation \cite{Mkhitaryan2014, Dobrovitski2009, Ma2016, Yang2008a} used previously to deal with square root of quantum operator).  
	As seen from Eq.~(\ref{eq: GenHam}), when the bath Hamiltonian $H_\mathrm{b}$ is neglected, the central spin experiences a static noise, i.e., a static distribution of $E_M$ due to the thermal distribution of the nuclear spin states. In this case, the central spin decoherence can be completely suppressed by spin echo.
	
	The bath interaction $H_\mathrm{b}$ (including both Zeeman energy and inter-nucleus spin coupling), which usually does not commute with the 
	operator $\Delta$, induces the dynamical fluctuation of the Overhauser field (as can be easily seen in the interaction picture $\hat{\Delta}(t)=e^{iH_\mathrm{b} t}\Delta e^{-iH_\mathrm{b} t}\ne \hat{\Delta}(0)$). This in turn will cause the dynamical fluctuation of the central spin splitting. To account for such dynamical fluctuations, which are usually much smaller than $E_M$, we separate 
	$\Delta$ into the mean-field part $\Delta_M$ (depending on the initial state of the bath) and the fluctuating part $\delta$, that is 
	\begin{equation}
	\Delta=\Delta_M+\delta.
	\end{equation}
	Up to the second order of the fluctuation part $\delta$, the resonance frequency is expanded as 
	\begin{equation}\label{eq:expand}
	\begin{aligned}
	E_M(\delta) 
	&=\sqrt{\varepsilon^2+(\Delta_M+\delta)^2}\\
	&\approx E_M+\frac{\Delta_M }{E_M}\delta + \frac{1}{2E_M}\delta^2.
	\end{aligned}
	\end{equation}
	In principle, the eigenstate wavefunctions of the central spin and hence the diagonal operator $\sigma_{\tilde{z}}^M$ also fluctuate with the Overhauser field. However, as illustrated in Appendix~\ref{sec: StaticBasis}, the wavefunction fluctuation induces much less decoherence effect than the frequency fluctuation in cases of interest in this paper. Therefore, we fix the diagonal operator $\sigma_{\tilde{z}}^M(\delta) $ to be $\sigma_{\tilde{z}}^M$, whose eigenstates $|\pm\rangle$ are determined with $\delta=0$.

	\begin{figure}[h!]
		\centering
		\includegraphics[width=0.8\linewidth]{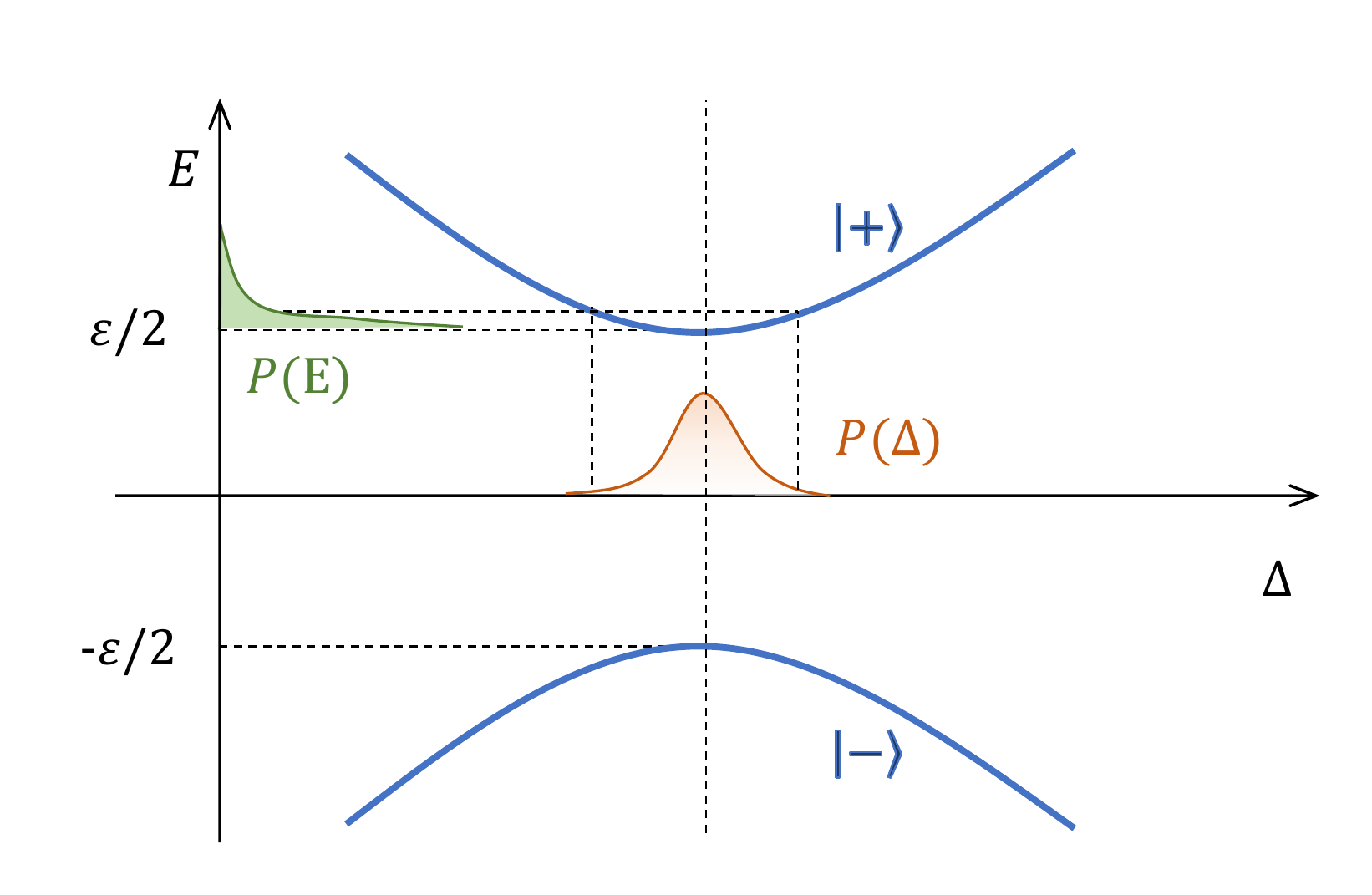}
		\caption{ Illustration of the clock transition. Near $\Delta = 0$, due to the parabolic dependence, the distribution of the transition frequency $P(E)$ is squeezed as compared with the distribution of the field $P(\Delta)$. Therefore, decoherence is suppressed at CT. }
		\label{fig:ct}
	\end{figure}
	
	As shown in Eq.~(\ref{eq:expand}) and Fig.~\ref{fig:ct}, the eigenenergy anti-crossing at $\Delta_M=0$ corresponds to the CT point. At the CT point, the energy is insensitive to the first order fluctuations, i.e., $dE_{M}/d\delta = 0$. Therefore CT has prolonged spin coherence times. 
	
	With the transformations and approximation, the Hamiltonian in Eq.~(\ref{eq: orgHam}) becomes
	\begin{equation}
	H = H_\mathrm{b} + \frac 12 \sum_M E_M(\delta) \lvert M \rangle\langle M\rvert \otimes\sigma^M_{\tilde z} 
	=H_\mathrm{b}+H_0.
	\end{equation}
	
	The coherence of the qubit for a certain bath state $\lvert M \rangle$ is characterized \cite{Yao2006, Liu2007} (for free-induction decay (FID), with extension to spin echo being straightforward) as 
	\begin{equation}
	L(t) = \langle M \rvert e^{iH^{(-)}t} e^{-iH^{(+)}t} \lvert M \rangle ,
	\end{equation}
	where the bath Hamiltonian conditioned on the central spin states $|\pm\rangle$ is
	\begin{equation}
	H^{\pm}\equiv H_\mathrm{b}\pm \frac{1}{2} E_M(\delta).
	\end{equation}
	Since $H_\mathrm{b}$ now can be regarded as a perturbation with respect to the unperturbed Hamiltonian $H_0$, the CCE theory \cite{Yang2008, Yang2009} is suitable for calculating $L(t)$. 
	
	In CCE, decoherence is expanded as the product of different orders of cluster correlations. For finite-time evolution, the expansion to a few orders will typically be sufficient. The $K$-order CCE (referred to CCE-$K$) is \cite{Zhao2012}
	\begin{equation}
	L^{(K)}(t) = \prod_{\lvert C\rvert \leq K} \tilde{L}_C(t),
	\end{equation}
	where $\lvert C\rvert$ is the cluster size, and the cluster correlation $\tilde{L}_C(t)$ is defined as
	\begin{equation}
	\tilde{L}_C(t) = \frac{L_C(t)}{\prod_{C^\prime \subset C}\tilde{L}_{C^\prime}},
	\end{equation}
	where $L_C(t)$ is the qubit decoherence caused by cluster $C$.  
	
	\section{Decoherence of NV center electron spin near CT}
	\label{sec: NV}
	We consider a negatively charged NV center \cite{Doherty20131, childress2007coherent}, whose electronic ground state is a triplet state with spin $S=1$. The NV center is coupled to $^{13}\mathrm{C}$ nuclear spins $\{\mathbf{I}_i\}$, which are spin-$1/2$'s with natural abundance $1.1\%$, randomly distributed on the diamond lattice. The Hamiltonian of the system \cite{Zhao2012} consists of three parts, 
	\begin{equation}
	H = H_{\mathrm{NV}} + H_\mathrm{b} + H_{\mathrm{int}},%
	\label{eq:FullHam}
	\end{equation}
	where $H_{\mathrm{NV}}$ and $H_\mathrm{b}$ are the Hamiltonians of NV center electron spin and the nuclear spins, respectively, and $H_{\mathrm{int}}$ is the interaction between the NV center and the bath. 
	
	In the very-weak field regime ($B\lesssim 0.1\,\mathrm{G}$), local strain-induced transverse anisotropy is not negligible, contributing a term to the NV center Hamiltonian in addition to the zero-field splitting energy and Zeeman energy of the central spin \cite{Ma2016},
	\begin{equation}
	H_{\mathrm{NV}} = D S_z^2 - \gamma_e BS_z + \varepsilon (S_x^2-S_y^2),
	\end{equation}
	where $S_z$ has the eigebasis $\{\lvert 0\rangle, \lvert\pm 1\rangle\}$, $D=2.87\,\mathrm{GHz}$ is the zero-field splitting, $\varepsilon$ is the strain-induced transverse anisotropy (a typical value $100\,\mathrm{kHz}$ is assumed in this paper according to experimental data \cite{Jamonneau2016}), $\gamma_e=-2.8\,\mathrm{MHz\,G^{-1}}$ is the gyromagnetic ratio of the electron spin, and $B$ is the external magnetic field along the NV axis. Note that a transverse field $B_{\perp}$ would only modify the energies as a second order perturbation $\sim \gamma_e^2B^2_{\perp}/D$, which is negligible in the weak field regime.
	
	The bath Hamiltonian consists of nuclear spin Zeeman energy and dipole-dipole interaction between nuclear spins
	\begin{equation}
	H_\mathrm{b} = -\gamma_n \sum_j \mathbf{B}\cdot \mathbf{I}_j + \sum_{i<j}\mathbf{I}_i\cdot\mathbb{D}_{ij}\cdot\mathbf{I}_j ,
	\end{equation}
	where $\gamma_n=1.1\,\mathrm{kHz\,G^{-1}}$ is the gyromagnetic ratio of $^{13}\mathrm{C}$ nuclear spins, and $\mathbb{D}_{ij}$ is the dipolar interaction tensor between the $i$th and $j$th nuclear spins \cite{Zhao2012}
	\begin{equation}
	\label{eq: dipole}
	\mathbb{D}_{ij} = \frac{\mu_0}{4\pi}\frac{\gamma_n^2}{r_{ij}^3}\left(\mathbb{I}-\frac{3\mathbf{r}_{ij}\mathbf{r}_{ij}}{r_{ij}^3}\right) ,
	\end{equation}
	where $\mu_0$ is the vacuum permeability and $\mathbf{r}_{ij}$ is displacement between the $i$th and $j$th nuclear spins. 
	
	In principle the nuclear spin Zeeman energy should include both the magnetic field along the NV axis and the transverse component. However, as will be made clear later in this paper, the amplitude and direction of the field are not essential as long as the nuclear Zeeman energy is weak (in comparison with hyperfine interaction) and does not commute with the hyperfine interaction. Without loss of generality, we assume the field is along the $z$ axis, i.e., ${\mathbf B}= B{\mathbf z}$.
	
	The electron and bath spins are coupled through the hyperfine interaction
	\begin{equation}
	H_{\mathrm{int}} = \mathbf{S}\cdot\sum_j\mathbb{A}_j\cdot\mathbf{I}_j \equiv {\mathbf S}\cdot {\mathbf h} ,
	\end{equation}
	where $\mathbb{A}_j$ is the hyperfine interaction tensor for the $j$th nuclear spin, and ${\mathbf h}$ denotes the Overhauser field.
	
	The transverse components of the Overhauser field ${\mathbf h}$ are neglected since their effects on the electron spin eigenenergies is mainly a second order perturbation $\sim h^2/D$, which is negligible ($\ll h$ and $\ll \varepsilon$). Thus 
	\begin{equation}
	H_{\rm int}\approx S_z\sum_j {\mathbf A}_j\cdot{\mathbf I}_j, 
	\end{equation}
	with 
	\begin{equation}
	\mathbf{A}_j = \mathbf{z}\cdot \frac{\mu_0}{4\pi}\frac{\gamma_e\gamma_n}{r_{j}^3}\left(\mathbb{I}-\frac{3\mathbf{r}_{j}\mathbf{r}_{j}}{r_{j}^3}\right) .%
	\label{eq: Knight}
	\end{equation}
	Here the Fermi contact part is neglected since the tightly bound orbits of the NV center 
	electrons have little overlap with the $^{13}$C nuclei.
	
	\subsection{NV center spin eigenstates under Overhauser field}
	
	One eigenstate of the Hamiltonian of Eq.~(\ref{eq:FullHam}) is $|0\rangle$. The other two eigenstates are formed by $|\pm 1\rangle$ as  
	\begin{equation}
	\left\{
	\begin{aligned}
	\lvert+\rangle &= \cos\frac{\theta_e}{2}\lvert+1\rangle + \sin\frac{\theta_e}{2}\lvert-1\rangle,\\
	\lvert-\rangle &= -\sin\frac{\theta_e}{2}\lvert+1\rangle + \cos\frac{\theta_e}{2}\lvert-1\rangle,
	\end{aligned}
	\right.%
	\label{eq:EleTrans}
	\end{equation}
	where $\theta_e = \tan^{-1}(\varepsilon/\omega_e)$ with $\omega_e = -\gamma_e B$ being the electron spin Larmor frequency. In the eigenbasis $\{\lvert 0\rangle,\lvert \pm\rangle\}$, the NV Hamiltonian is diagonalized as $H_{\mathrm{NV}} = \sqrt{\varepsilon^2 + \omega_e^2}S_{\tilde{z}}$ ($D$ dropped hereafter). 
	
	Taking into account the Overhauser field along the $z$-axis, $h_z = \sum_j \mathbf{A}_j\cdot \mathbf{I}_j$, the NV center Hamiltonian becomes
	\begin{equation}
	H_e =  D S_z^2 + (\omega_e + h_z) S_z + \varepsilon (S_x^2-S_y^2),
	\end{equation} 
	which is formally diagonalized as 
	\begin{equation}
	\label{eq: H_e}
	H'_e = \sqrt{\varepsilon^2 + (\omega_e+h_z)^2}S'_{\tilde{z}}. 
	\end{equation}
	
	
	If Eq.~(\ref{eq: H_e}) is expanded to the second order of $h_z$, the eigenenergy contains terms $\propto {\mathbf I}_i \cdot {\mathbf A}_i{\mathbf A}_j\cdot {\mathbf I}_j$, which are effective long-range interactions among nuclear spins. Such long range interactions depend on the distances from nuclear spins to the electron spin rather than the relative distances between nuclear spins. For relatively close nuclear spins (for instance within $\sim 1\,\mathrm{nm}$ from the electron spin), the hyperfine-mediated interaction ($\sim \mathrm{kHz}$) will be more important than their intrinsic dipole-dipole interaction given in Eq.~(\ref{eq: dipole}) ($\sim 0.1\,\mathrm{kHz}$).  Therefore, a better choice of the working basis is the eigenstates of the hyperfine interaction, denoted as  $\{\lvert \uparrow\rangle_j, \lvert \downarrow\rangle_j\}$ for the $j$th nuclear spin. 
	In so-chosen basis, the Hamiltonian of Eq.~(\ref{eq:FullHam}) can be recast as
	\begin{equation}
	H=\sum_M \sqrt{\varepsilon^2+(\omega_e+h_z^M)^2}\lvert M\rangle\langle M\rvert\otimes S_{\tilde{z}}^{M} + H_\mathrm{b},%
	\label{eq:NewFullHam}
	\end{equation}
	where $h_z^M= \langle M\rvert h_z\lvert M\rangle$ is the Overhauser field for the bath eigenstates $|M \rangle\equiv \otimes_{j}|m_j\rangle_j$ with $m_j=\uparrow$ or $\downarrow$. In general, $S_{\tilde{z}}^{M}$ depends on the Overhauser field, which can be fluctuating in time. However, as shown in Appendix~\ref{sec: StaticBasis}, due to the fact that the Overhauser field fluctuation is small and slow, we can take $S_{\tilde{z}}^M$ as fixed and independent of the fluctuation $\delta$. Note that unlike in previous works \cite{Mkhitaryan2014, Dobrovitski2009, Ma2016, Yang2008a}, the hyperfine interaction is treated beyond the second-order approximation.
	
	In the following, we assume that the central spin is initialized as $\lvert\psi_e(0)\rangle=1/\sqrt{2}(\lvert+\rangle+\lvert0\rangle)$. For temperature much higher than the nuclear spin eigenenergies ($\mathrm{MHz} \sim \mu\mathrm{K}$), the nuclear spins can be described by an identity density matrix, $\rho=2^{-N}$. The conditional Hamiltonian for central spin at $\lvert+\rangle$ and $\lvert0\rangle$ states are
	\begin{subequations}\label{eq:subHam}
	\begin{align}
	H^{(+)} &= \sum_M E_M\lvert M\rangle\langle M\rvert + H_\mathrm{b} , \\
	H^{(0)} &=  H_\mathrm{b}. 
	\end{align}
	\end{subequations}
	
	\subsection{Cluster-correlation Expansion}
	\label{SubSec: CCE}

	\begin{figure*}[ht]
		\includegraphics[width=\linewidth]{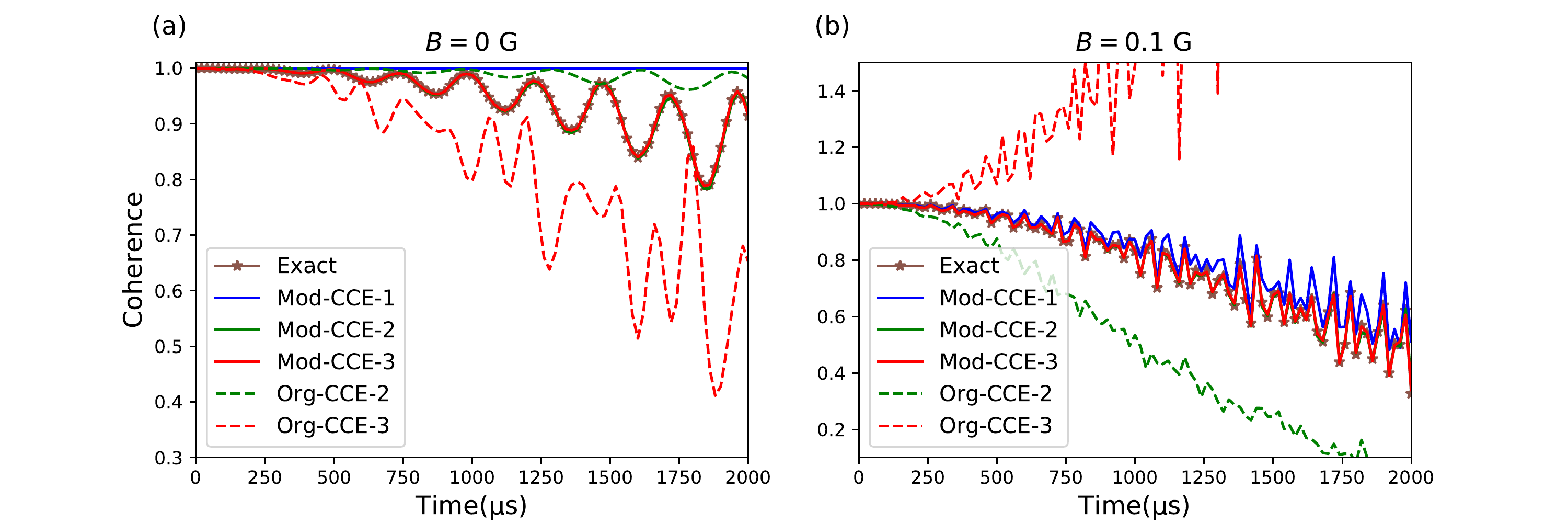}
		\caption{\label{fig:comparecce} Comparison of  different orders of the modified CCE and the original CCE to exact numerical results for a random bath state near the clock transition. The number of nuclear spins is $N=10$. (a) Hahn echo result with $B=0\,\mathrm{G}$, (b) Hahn echo result with $B=0.1\,\mathrm{G}$. The results show that the modified CCE has converged to the exact result with truncation order of 2. While the original CCE does not converge. }
	\end{figure*}
	
	When the number of nuclear spins taken into account is small ($N\sim 20$), the spin decoherene problem defined by Hamiltonian of Eq.~(\ref{eq:NewFullHam}) can be exactly solved by the numerical diagonalization or time integration. However, typically a much larger number of nuclear spins ($N>100$) are needed to obtain converged results for calculating the central spin decoherence in a lattice. Now by treating the hyperfine interaction non-perturbatively, the CCE can be readily applied to central spin decoherence near CTs. We'll show that CCE-2 already gives converged results.
	
	For a pure bath initial state $\lvert M \rangle$, the coherence of the electron spin expanded up to CCE-2 is
	\begin{equation}
	L^M(t) \approx e^{iE_M t}\prod_j \tilde{L}_j^M \prod_{i<j} \tilde{L}_{ij}^M,
	\end{equation}
	where the phase factor $e^{iE_M t}$ is due to the empty-cluster correlation (the bath Hamiltonian is neglected), $\tilde{L}_j^M$ is the single-spin cluster correlation from the $j$th nuclear spin, and $\tilde{L}_{ij}^M$ is the two-spin (pair) correlation from the $i$th and $j$th nuclear spins. 
	
	Because the hyperfine interaction in general does not commute with the nuclear Zeeman energy, i.e., the $z$ axis is different from the hyperfine quantization axis, the Zeeman terms can induce transitions between different hyperfine eigenstates $|M\rangle$, causing fluctuations of the Overhauser field $h_z^M$, and in turn fluctuation of the eigenenergy of the electron spin states. Therefore, the single spin dynamics (CCE-1) contributes significantly to the decoherence of NV center spin under low field, which is in contrast to the case of relatively strong field where the processes not conserving the Zeeman energy  are suppressed. 
	
	For FID, the electron spin decoherence caused by single-spin dynamics is given by
	\begin{equation}
	\tilde{L}_j^M = \frac{1}{e^{iE_M t}}\langle M\vert e^{i H_j^{(+)}t} e^{-i H_j^{(0)}t}\vert M\rangle, 
	\end{equation}
	with the Hamiltonian for the $j$th spin,
	\begin{subequations}
	\begin{align}
	H_j^{(+)} &= \sum_{m_j} E_j \lvert m_j\rangle\langle m_j\rvert + h_j^{(0)}, \\
	H_j^{(0)} &= -\gamma_n \mathbf{B}\cdot \mathbf{I}_j.
	\end{align}
	\end{subequations}
	
	The electron spin decoherence caused by the nuclear spin pair is given by
	\begin{equation}
	\tilde{L}_{ij}^M = \frac{1}{e^{iE_M t}\tilde{L}_i^M\tilde{L}_j^M}\langle M\vert e^{i H_{ij}^{(+)}t} e^{-i H_{ij}^{(0)}t}\vert M\rangle,
	\end{equation}
	with the Hamiltonian for the $i$th and $j$th nuclear spin pair,
	\begin{subequations}
	\begin{align}
	H_{ij}^{(+)} &= \sum_{m_i m_j}E_{ij}\lvert m_i m_j\rangle\langle m_i m_j\rvert + H_{ij}^{(0)}, \\
	H_{ij}^{(0)} &= - \gamma_n \mathbf{B}\cdot \mathbf{I}_i - \gamma_n \mathbf{B}\cdot \mathbf{I}_j + \mathbf{I}_i \cdot\mathbb{D}_{ij}\cdot\mathbf{I}_j.
	\end{align}
	\end{subequations}
	
	For Hahn echo, electron spin decoherence caused by the single and pair correlations becomes
	\begin{equation}
	\left\{
	\begin{aligned}
	\tilde{L}_j^M &= \langle M\vert e^{i H_j^{(+)}t} e^{i H_j^{(0)}t} e^{-i H_j^{(+)}t} e^{-i H_j^{(0)}t}\vert M\rangle, \\
	\tilde{L}_{ij}^M &= \frac{1}{\tilde{L}_i^M\tilde{L}_j^M} \langle M\vert e^{i H_{ij}^{(+)}t} e^{i M_{ij}^{(0)}t} e^{-i H_{ij}^{(+)}t} e^{-i H_{ij}^{(0)}t}\vert M\rangle.
	\end{aligned}
	\right.
	\end{equation}
	
	To verify the correctness and convergence of the CCE, we compare the CCE results with the exact numerical result. In exact numerical simulation, we integrate the Schr\"odinger equation using a standard numerical algorithm (leap-frog), $\lvert \psi(t+\Delta t)\rangle \approx \lvert \psi(t-\Delta t)\rangle - 2i\Delta t H\lvert \psi(t)\rangle$, with the stability criterion that the time step $\Delta t<1/||H||$. Figure~\ref{fig:comparecce} shows the comparison for a bath of 10 nuclear spins in a state $\lvert M\rangle$ with vanishing Overhauser field ($ h_z^M \approx 0$). At least for the time scale less than $2\,\mathrm{ms}$ (which is much greater than the decoherence time) the modified CCE-2 converges to the exact result. However, the original CCE, which contains mediated long-range interactions, does not converge. Moreover, unlike the pair-correlation-dominated electron spin decoherence for the NV center in high field regime \cite{Zhao2012, Yang2016} or double QDs \cite{Yang2008a} in Appendix~\ref{Sec: Double-QDs}, the single spin correlation (CCE-1) dominates the decoherence in the weak magnetic field regime. 
	
	\subsection{Results and Discussions}
	\label{SubSec: Res}
	
	Since the decoherence for FID is just a result of inhomogeneous broadening \cite{Yang2016}, which does not need CCE for calculation, in this subsection, we are interested in studying the decoherence for Hahn echo and the dependence of $T_2$ on external magnetic fields.

	\begin{figure}[h!]
		\centering
		\includegraphics[width=\linewidth]{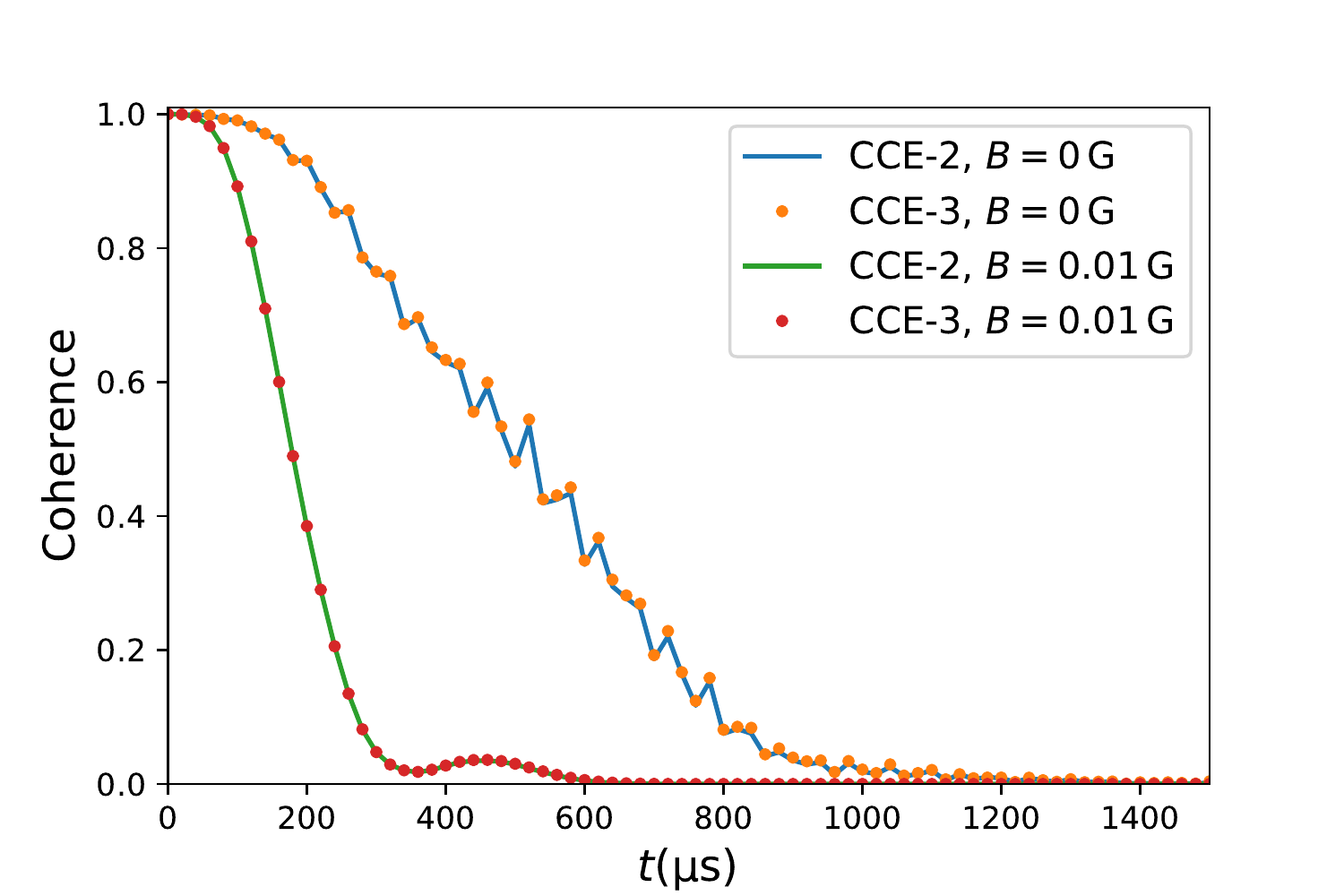}
		\caption{Comparison between CCE-2 and CCE-3 results at and away from the clock transition for a pure bath state with vanishing Overhauser field. In the randomly chosen spatial configurations of $^{13}$C nuclei on the lattice, the nearest nuclear spin is $0.77\,\mathrm{nm}$ ($A_1=60.2\,\mathrm{kHz}$) from the central electron spin. The number of nuclear spins included in the calculation is $N=500$. }
		\label{fig:comparecce23}
	\end{figure}
	
	For Hahn echo, the static thermal noise is eliminated, and the electron spin decoherence is caused by the dynamical fluctuations of the nuclear spin bath. In Fig.~\ref{fig:comparecce23}, we compare the results from different orders of CCE for a nuclear spin bath ($N=500$) in a pure state $|M\rangle$ with vanishing Overhauser field ($h_z^M\approx 0$). The comparison shows that, after absorbing the long-range interactions as eigenenergy fluctuations, the modified CCE converges even at CT ($B=0$). 
	
\begin{figure}[h]
		\includegraphics[width=\linewidth]{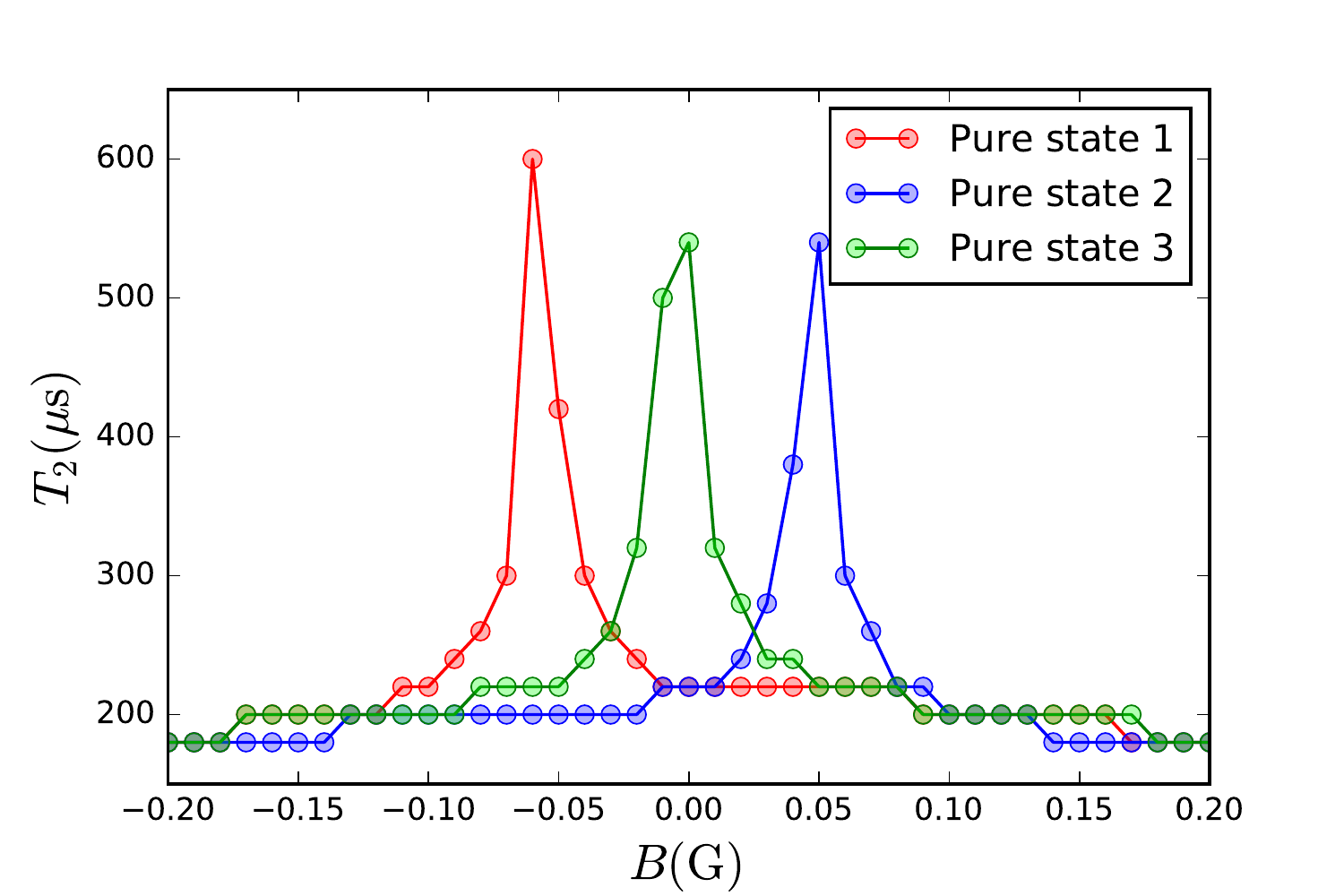}
		\caption{\label{fig:T2time3Pure} Dependence on external field of NV center electron spin coherence time of Hahn echo for three different pure bath states. The results are calculated by the modified CCE-2. The number of nuclear spins included in the calculation is $N=500$. The maximum of the coherence time occurs where external magnetic field cancels the Overhauser field.}
	\end{figure}

	Figure~\ref{fig:T2time3Pure} shows the coherence time of Hahn echo for three different initial bath states. The dependence presents sharp peaks with the maximum coherence times appear  where external magnetic field approximately cancels the Overhauser field for that specific bath state (i.e., $\omega_{\rm e}+h_z^M=0$). Such peak features demonstrate that the coherence of electron spin is well protected at the clock transition. Also, the coherence time as a function of the magnetic field has similar line width for different pure initial states, since the dynamical fluctuations of the Overhauser field is almost independent of the initial state $|M \rangle$.  
	
	In order to get the ensemble-averaged result, in principle, we need to take average of $L(t)$ over all bath states. However, since the remote nuclear spins have negligible effects on the static part of Overhauser field, which determines the location of CTs, we can set the remote nuclear spins at randomly chosen initial states and only take the average over all states of several close nuclear spins ($\sim 10$). 
	
	\begin{figure}[h]
		\includegraphics[width=\linewidth]{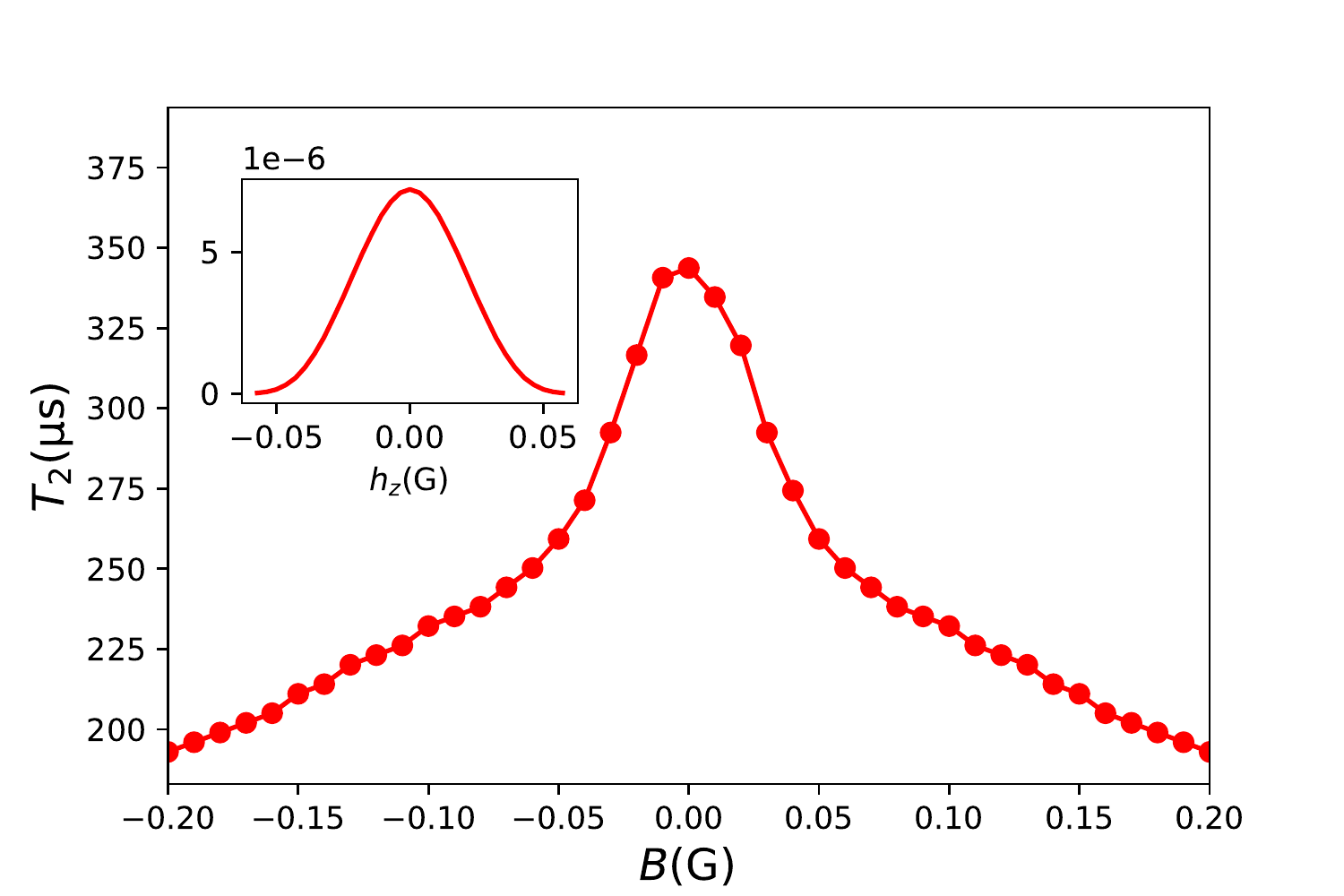}
		\caption{\label{fig:T2EchoMix} Ensemble-averaged result of NV electron spin coherence time for Hahn echo. The number of nuclear spins is $N=200$. The coherence time is obtained from the decoherence profile averaged over all states of the nearest 10 bath spins. Inset: The distribution of Overhauser field.}
	\end{figure}
	
	Figure~\ref{fig:T2EchoMix} shows the ensemble-averaged result of $T_2$ as a function of the magnetic field. Considering the averaging is expensive in computation, we have reduced the number of nuclear spins to $N=200$. The nearest $10$ nuclear spin states are fully averaged (totally $2^{10}=1024$ pure states) with remote nuclear spins at random pure states. We can see that the sharp line feature (Fig.~\ref{fig:T2time3Pure}) for the pure bath state case has been lost in the ensemble-averaged result.

	\section{Conclusion}
	\label{sec: conclusion}
	In this paper, we theoretically investigated a class of decoherence problems which can be solved by the modified CCE method. As an example, we study in details the NV center decoherence in the near-zero magnetic field.
	By transforming the basis, the hyperfine mediated bath interaction is absorbed as fluctuation of the central spin eigenenergies. We find that the dynamical fluctuation of the Overhauser field has a much larger effects on the central spin eigenenergies than on the wavefunctions of the eigenstates, so the static approximation can be adopted for the central spin eigenstates. In the modified CCE, the static part of the Overhauser field is treated exactly without resorting to the second order expansion method. For numercial simulations, we have calculated the NV center decoherence in the near zero magnetic field regime, which displays quite different properties than in the high field regime. We also solve the single-triplet decoherence problem and show that CCE-2 also converges for this system in Appendix~\ref{Sec: Double-QDs}. One more example, namely, the decay of Rabi oscillation, is formulated in Appendix~\ref{sec: RotaryEcho}. 
	
	\begin{acknowledgments}
	This work was supported by Hong Kong Research Grants Council ANR/RGC Joint Research Scheme Project A-CUHK403/15.
	\end{acknowledgments}
	
	\appendix
	
	\section{Static Approximation of central spin eigenstates}
	\label{sec: StaticBasis}
	
	In most cases of electron spin decoherence in nuclear spin baths, the dynamical part of the fluctuation ($\delta\equiv h_z(t)-\langle M|h_z|M\rangle$) is much smaller than the electron spin splitting at the CT, i.e., $\delta\ll \varepsilon$, and the fluctuation of $\tilde{\delta}(t)=e^{iH_\mathrm{b}t}\delta e^{-iH_\mathrm{b}t}$ is much slower than the electron spin dynamics, for $|H_\mathrm{b}|\ll \varepsilon$. Therefore, the basis of $\sigma_{\tilde{z}}^M(\delta)$ adiabatically follows the direction of $(\Delta_M+\tilde{\delta}(t), 0, \varepsilon)$. 
		
	Starting from the initial state $\lvert \psi(0)\rangle = \frac{1}{\sqrt{2}} (\lvert +\rangle + \lvert -\rangle) \lvert J\rangle$, the state at time $t$ 
	\begin{equation}
	\begin{aligned}
	\lvert \psi(t)\rangle 
	&\approx \frac{1}{\sqrt{2}} e^{-iH_\mathrm{b} t} \mathcal{T} e^{-i\int_0^t \sqrt{\varepsilon^2 + (\Delta_M + \tilde{\delta}(\tau))^2} d\tau} \lvert +', M\rangle \\
	&\quad + \frac{1}{\sqrt{2}} e^{-iH_\mathrm{b} t} \mathcal{T} e^{i\int_0^t \sqrt{\varepsilon^2 + (\Delta_M + \tilde{\delta}(\tau))^2} d\tau} \lvert -', M\rangle,
	\end{aligned} 
	\end{equation}
	where $\lvert \pm'\rangle$ are the instantaneous eigenstates in direction $(\Delta_M+\tilde{\delta}(t), 0, \varepsilon)$. 
	
	The coherence is
	\begin{equation}
	\label{eq: Ltad}
	\begin{aligned}
	L(t) &= \left\langle \left(\sigma_x+i\sigma_y\right)\right\rangle/2 \\
	&\approx \frac12 \langle M\rvert \bar{\mathcal{T}}e^{i\int_0^t \sqrt{\varepsilon^2 + (\Delta_M + \tilde{\delta}(\tau))^2} d\tau}\\
	&\qquad \mathcal{T} e^{-i\int_0^t \sqrt{\varepsilon^2 + (\Delta_M + \tilde{\delta}(\tau))^2} d\tau} \lvert M\rangle,
	\end{aligned}
	\end{equation}
	where $\bar{\mathcal{T}}$ is the anti-time-ordering operator and we have used that $\langle n\vert m\rangle = \delta_{m,n}$ with $m,n=\pm'$. The results can be immediately identified as being equivalent to the approximation $\lvert \pm'\rangle \approx \lvert \pm \rangle$ during the evolution. 
		
	Therefore, we justify the approximation
	\begin{equation}
	\sigma_{\tilde z}^M(\delta)\approx \sigma^M_{\tilde z}.
	\end{equation}

	\section{Decoherence of GaAS Double Quantum dot}
	\label{Sec: Double-QDs}
	Here we consider a gate-defined symmetric GaAs double quantum dot \cite{Burkard1999, Coish2005, Yang2008a} with one electron in each dot. In the zincblende structure $\mathrm{As}$ and $\mathrm{Ga}$ are located in two interpenetrating face-centered cubic lattices. All the isotopes have a spin quantum number of $3/2$, and the natural abundances for ${}^{75}\mathrm{As}$, ${}^{69}\mathrm{Ga}$ and ${}^{71}\mathrm{Ga}$ are shown in Table~\ref{tab: material} \cite{Taylor2007, Medford2012}. Assuming an external magnetic field is applied along [001] direction, which is also taken to be the $z$ axis in this section, the Hamiltonian for the electron-nuclear system is
	\begin{align}\label{}
	H=\omega_{e}(S_1^z+S_2^z)+J_{\mathrm{ex}}\mathbf{S}_1\cdot\mathbf{S}_2+S_1^z h_1^z+S_2^z h_2^z+H_{\mathrm{b}},
	\end{align}
	where $\mathbf{S}_1$ and $\mathbf{S}_2$ are the electron spin operators, $\omega_e = -\gamma_e^\ast B$ is the electron Zeeman splitting with $\gamma_e^\ast = 1.32\times 10^{11}\,\mathrm{S^{-1}\,T^{-1}}$ being the effective gyromagnetic ratio for GaAs dot\cite{Yao2006, Liu2007, Yafet1961}, $J_{\mathrm{ex}}$ (taken to be $-1\,\mathrm{\mu ev} \approx -0.24\,\mathrm{GHz}$ in calculation) is the exchange interaction due to inter-dot tunneling \cite{Burkard1999, Petta2005}, and ${h}_k^z=\sum_j {a}_{k,j} {I}_{k,j}^z$ ($k=1,2$) is the Overhauster field with ${a}_{k,j}$ being the Fermi contact hyperfine coefficient for the $j$th nuclear spin in the $k$th dot. A large external magnetic field is applied ($B \geq 1\,\mathrm{T}$) and the electron-nuclear flip-flop processes are suppressed due to  large Zeeman splitting mismatch, so the off-diagonal hyperfine interaction is neglected. Here the total bath Hamiltonian is
	\begin{equation}
	H_\mathrm{b} = H_{\mathrm{b},1} + H_{\mathrm{b},2},
	\end{equation}
	where $H_{\mathrm{b},k}$ (with $k=1,2$) is the bath Hamiltonian for the $k$th dot. $H_{\mathrm{b},k}$ is assumed to be independent for the two dots with the form 
	\begin{equation}
	\begin{aligned}
	H_{\mathrm{b},k} &= \sum_{k,j} \omega_{k,j} I_{k,j}^z + \sum_{k, i<j} \mathbf{I}_{k,i} \cdot \mathbb{D}_{k, i;k,j}\cdot\mathbf{I}_{k, j} \\
	&\approx \sum_{k,j} \omega_{k,j} I_{k,j}^z + {\sum_{k, i<j}}^{\prime} D_{k, i;k, j} I_{k,i}^z I_{k,j}^z \\
	&\quad + {\sum_{k,i<j}}'' \frac{D_{k, i;k, j}}{2} (3I_{k,i}^z I_{k,j}^z - \mathbf{I}_{k,i} \cdot \mathbf{I}_{k,j}) ,
	\end{aligned}
	\end{equation}
	where $D_{k, i;k, j} = \frac{\mu_0\gamma_{k,i} \gamma_{k,j}}{4\pi r_{k, i;k, j}^3} (1-3\cos^2 \theta_{k, i;k, j})$ is the dipolar hyperfine coefficient in the high field regime, the summation with a prime runs over only the hetero-nuclear pairs and summation with a double prime runs over only the homo-nuclear pairs. Due to the large external magnetic field ($\omega_n \gg D_{ij}$), hetero-nuclear pair flip-flops are suppressed, so only the secular terms which conserve the Zeeman energy have been kept \cite{Yang2008a}. 
	
	The Fermi contact hyperfine coefficient is $a_{k,j} = A_n a^3 \lvert f(\mathbf{r}_{k,j}) \rvert^2$, with $a = 5.63\,\text{\AA}$ the lattice constant. $A_n = \frac 23 \mu_0\gamma_e \gamma_n d_n$ is the hyperfine constant for the $n$th isotope, with $\gamma_e = -1.76\times 10^{11}\,\mathrm{S^{-1}\, T^{-1}}$ being the free electron gyromagnetic ratio, $\gamma_n$ the nuclear gyromagnetic ratio, and $d_n$ the electronic density \cite{Klauser2008}. These parameters are shown in Table~\ref{tab: material}. $f(\mathbf{r})$ is the envelope wavefunction of the ground state. The confinement of GaAs dot along the $z$ direction and the lateral directions is assumed to be hard-wall and parabolic \cite{Yao2006}, respectively. Therefore, the envelope function 
	\begin{equation}
	f(\mathbf{r}) = \sqrt{\frac{2}{L_z}} \cos\frac{\pi z}{L_z} \cdot \frac{1}{\sqrt{\pi}\rho_0} e^{-\frac{\rho^2}{2\rho_0^2}}, \lvert z\rvert \leq \frac{L_z}{2}, 
	\end{equation}
	where $L_z$ is the dimension in $z$ direction, and $\rho_0$ is the Fock-Darwin radius \cite{Kouwenhoven2001, Jacak1998}. In the calculation, $L_z = 6\,\mathrm{nm}$ and $\rho_0 = 30\,\mathrm{nm}$. 
	
	\begin{table}[h]
		\caption{\label{tab: material}%
			Parameters for nuclear spins in GaAs.
		}
		\begin{ruledtabular}
			\begin{tabular}[b]{lccc}
				{} & ${}^{75}\mathrm{As}$ & ${}^{69}\mathrm{Ga}$ & ${}^{71}\mathrm{Ga}$\\
				\hline
				Abundance & $100\%$ & $60.1\%$ & $39.9\%$ \\
				$\gamma_n (\times 10^7\,\mathrm{S^{-1}\, T^{-1}})$ & 4.60 & 6.44 & 8.18 \\
				$d_n (\times 10^{25}\,\mathrm{cm^{-3}})$ & 9.8 & 5.8 & 5.8  \\
				$A_n (\times 10^9\,\mathrm{S^{-1}})$ & 69.8 & 58.1 & 73.8 \\
			\end{tabular}
		\end{ruledtabular}
	\end{table}

	The qubit for the electron spin is formed by the singlet state $|S\rangle=(|\uparrow\downarrow\rangle-|\downarrow\uparrow\rangle)/\sqrt{2}$ and the unpolarized triplet state $|T_0\rangle=(|\uparrow\downarrow\rangle+|\downarrow\uparrow\rangle)/\sqrt{2}$ (with the two polarized triplet state $|T_+\rangle=|\uparrow\uparrow\rangle$ and $|T_-\rangle=|\downarrow\downarrow\rangle$ largely decoupled from the qubit space). In the singlet-triplet basis, the total Hamiltonian is \cite{Yang2008a}
	\begin{align}\label{}
	H=J_{\mathrm{ex}} S_z+ (h_1^z-h_2^z) S_x+H_{\mathrm{b}},
	\end{align}
	where $S_z=(|T_0\rangle\langle T_0|-|S\rangle\langle S|)/2$ and $S_x=(|T_0\rangle\langle S|+|S \rangle\langle T_0|)/2$. Then the pure-dephasing Hamiltonian in the central qubit eigenstate basis is
	\begin{align}\label{}
	H=\sqrt{J_{\mathrm{ex}}^2+(h_1^z-h_2^z)^2}S_{\tilde{z}}+H_{\mathrm{b}},
	\end{align}
	where we adopt the static approximation for $S_{\tilde z}$ as demonstrated in Appendix~\ref{sec: StaticBasis}. 
	
	To verify the convergence of CCE, we only consider single sample results for a pure initial state $\lvert M \rangle$ generated by setting $m_j$ for each nuclear spin as one of values $\{-3/2, -1/2, 1/2, 3/2\}$ randomly with the same probability, in the light of the high-temperature condition for nuclear spins ($k_{\mathrm{B}}T(10\,\mathrm{mK}) \sim 0.2\,\mathrm{GHz} \gg \omega_{k,j}$). The positions of nuclear spins in the bath are randomly placed according to their respective natural abundances (Table~\ref{tab: material}). It should be noted that unlike NV center case, CCE-1 Hamiltonians $H_j^{\pm} = \pm \frac 12\sqrt{J_{\mathrm{ex}}^2+(h_1^z-h_2^z)^2} + \omega_j I_j^z$ commute, making no contribution to the decoherence of the central qubit. 
	
	\begin{figure}[h!]
		\includegraphics[width=0.8\linewidth]{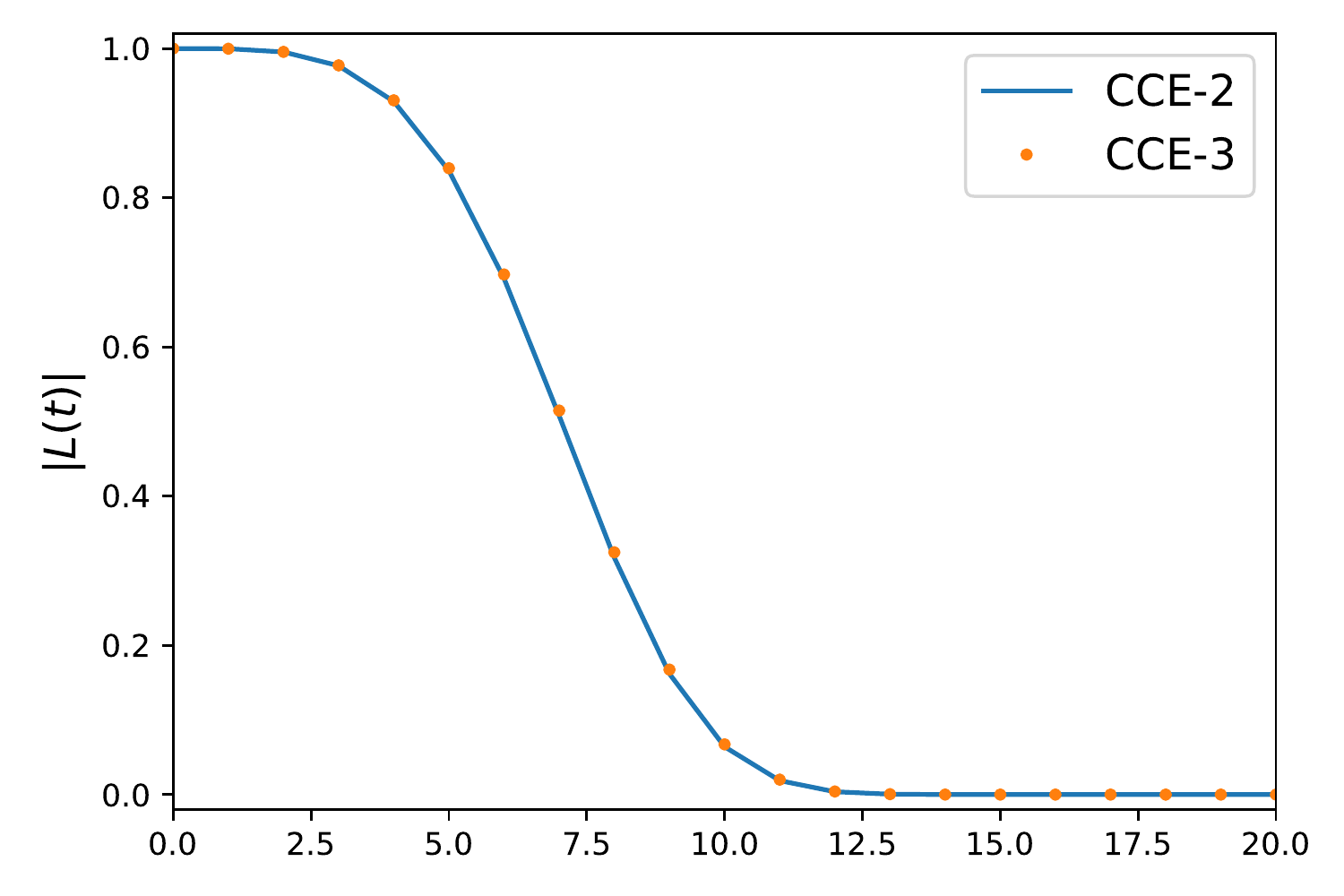}
		\caption{\label{fig: CCE23_Echo} Comparison of CCE-2 and CCE-3 results for singlet-triplet decoherence with Hahn echo applied in a GaAs double quantum dot. The initial state of the nuclear spin bath is randomly chosen as  $\lvert M \rangle$. The number of nuclear spins $N=1.6\times 10^6$. }
	\end{figure}
	
	Figure~\ref{fig: CCE23_Echo} shows the results from CCE-2 and CCE-3, which verifies that CCE-2 also gives the converged result in this system.

	\section{Decay of Rabi Oscillation and Rotary Echo}
	\label{sec: RotaryEcho}
	In this section we show that decay of Rabi oscillations \cite{Dobrovitski2009}, where interesting power law decay and phase shift of rotations were reported previously \cite{Koppens2007, Coish2005}, and rotary echos \cite{Laraoui2011, Mkhitaryan2014} also belong to the general class of problems that can be solved using the modified CCE. Suppose the central spin is under a large static magnetic field in the $z$ axis (hence the pure dephasing model) and a strong Rabi driving field in the $x$ direction, the Hamiltonian is \cite{Solomon1959, Laraoui2011}
	\begin{align}\label{}
	H=\varepsilon S_x+ (\Delta+h_z) S_z+H_{\mathrm{b}},
	\end{align}
	where $\varepsilon$ is the Rabi frequency in rotating frame, $\Delta=\omega_e-\omega$ is the detuning between the Zeeman frequency of the central spin and the frequency of driving field, and $h_z=\sum_iA_iI_i^z$ the Overhauster field from the nuclear spin bath. The pure-dephasing Hamiltonian in the central spin eigenstate basis is
	\begin{align}\label{}
	H=\sqrt{\varepsilon^2+(\Delta+h_z)^2} S_{\tilde{z}} + H_\mathrm{b}.
	\end{align}
	
	This Hamiltonian is immediately recognized as belonging to the class discussed in the main text, and therefore solvable using the modified CCE. 
	
	Rotary echoes (echoes in rotating frame) are produced by reversing the phase of the driving field at times $\{t_{1},t_{2}\cdots{t_{N}}\}$ for the system evolution from 0 to $t$, so that the total Hamiltonian becomes time-dependent as
	\begin{align}\label{}
	H_N(t)=f(t) \varepsilon S_x+ (\Delta+h_z) S_z+H_\mathrm{b},
	\end{align}
	where $f(t)=(-1)^{k}$ for $[t_{k},t_{k+1}]$ ($k=0,1,\cdots,N$ with $t_0=0$ and $t_{N+1}=t$). 
	
	The rotary echo signal can be calculated with the modified CCE in a similar way to the calculation of the Hahn echo at the clock transitions.
	
	\bibliography{Reference}
\end{document}